\documentclass[prr,twocolumn,showpacs,preprintnumbers,amsmath,amssymb,superscriptaddress]{revtex4-1}
\usepackage{amsmath,amssymb,amsfonts} 	% Typical maths resource package
\usepackage{graphicx}
\usepackage{subfigure}

\renewcommand{\[}{\begin{equation}}
\renewcommand{\]}{\end{equation}}
\def\bea{\begin{eqnarray}}
\def\eea{\end{eqnarray}}
\def\nn{\nonumber\\}

\newcommand{\intrc}{\int_{\rm cell} d{\bf r} \;}
\newcommand{\emi}[1]{{\rm e}^{-i #1}}
\newcommand{\ei}[1]{{\rm e}^{i #1}}
\newcommand{\B}{{\bf B}}

\newcommand{\intk}{\int_{\rm BZ} \frac{ d{\bf k}}{(2\pi)^d} \;}

\newcommand{\M}{{\bf M}}

\newcommand{\CQ}{{\cal Q}}
\newcommand{\CP}{{\cal P}}
\newcommand{\CX}{{\cal X}}

\newcommand{\p}{{\bf p}}
\newcommand{\q}{{\bf k}}
\renewcommand{\k}{{\bf k}}

\newcommand{\vc}{V_{\rm cell}}
\newcommand{\CH}{{\cal H}}
\newcommand{\CA}{{\cal A}}
\newcommand{\CB}{{\cal B}}
\newcommand{\CC}{{\cal C}}
\newcommand{\dkk}{\partial_{{\bf k}}}

\newcommand{\kk}{\mbox{\boldmath$\kappa$}}
\renewcommand{\r}{{\bf r}}
\newcommand{\R}{{\bf R}}

\newcommand{\da}{\partial_{k_\alpha}\!}
\newcommand{\db}{\partial_{k_\beta}\!}

\newcommand{\equ}[1]{Eq.~(\ref{#1})}
\newcommand{\eqs}[2]{Eqs.~(\ref{#1}) and (\ref{#2})}

\def\bra#1{\langle#1\vert}
\def\ket#1{\vert#1\rangle}
\def\ev#1{\langle#1\rangle}
\def\me#1#2#3{\langle#1| \, #2 \, |#3\rangle}
\def\runtime{(\the\time)\qquad\the\month/\the\day/\the\year}% get current time
\def\today
 {\count10=\year\advance\count10 by -2000 \number\day--\ifcase
  \month \or Jan\or Feb\or Mar\or Apr\or May\or Jun\or
             Jul\or Aug\or Sep\or Oct\or Nov\or Dec\fi--\number\count10}

\def\hour{\count10=\time\count11=\count10
\divide\count10 by 60 \count12=\count10
\multiply\count12 by 60 \advance\count11 by -\count12\count12=0
\number\count10 :\ifnum\count11 < 10 \number\count12\fi\number\count11}

\begin{document}

%%%%%%%%%%%%%%%%%%%%%%%%%%%%%%%%%%%%%%%%%
\title{X-ray circular dichroism versus orbital magnetization}

\author{Raffaele Resta}
\email{resta@iom.cnr.it}
\affiliation{Istituto Officina dei Materiali IOM-CNR, Strada Costiera 11, 34151 Trieste, Italy}
\affiliation{Donostia International Physics Center, 20018 San Sebasti{\'a}n, Spain}

\date{\today}

\begin{abstract} The x-ray magnetic circular dichroism (XMCD) sum rule yields an extremely useful ground-state observable, which provides a quantitative measure of spontaneous time-reversal symmetry breaking (T-breaking) in a given material. I derive here its explicit expression within band-structure theory, in the general case: trivial insulators, topological insulators, and metals. Orbital magnetization provides a different measure of T-breaking in the electronic ground state. The two observables belong to the class of ``geometrical'' observables; both are local and admit a ``density'' in coordinate space. In both of them one could include/exclude selected groups of bands, in order to acquire element-specific information about the T-breaking material. Only in the case of an isolated flat band the contributions to the two observables coincide. Finally, I provide the corresponding geometrical formula---in a different Hilbert space---for a many-body interacting system.
\end{abstract}

\maketitle

\section{Introduction}

Since the very popular 1992 paper by Thole et al. \cite{Thole92}, x-ray circular dichroism (XMCD) has been widely regarded as an approximate probe of orbital magnetization in bulk solids. Some years later it was clearly recognised that the XMCD sum rule $I_{\alpha\beta}$ (defined below) provides insight into magnetic properties of solids, although such ``magnetic properties'' do not coincide with orbital magnetization except in the extreme atomic limit \cite{Kunes00}. It must be mentioned that at the time no sound theory of orbital magnetization in bulk solids was available. Orbital magnetization $\M$ is by definition the derivative of the free-energy density with respect to magnetic field (orbital term thereof, and with a minus sign). After Ref. \cite{Kunes00} it is then pretty clear that $I_{\alpha\beta}$ is the free-energy derivative with respect to a different T-breaking probe: circularly polarized light, integrated over the whole spectrum. Owing to a fluctuation-dissipation theorem, a frequency-integrated dynamical probe becomes effectively a static one; the said probe has the virtue of coupling to orbital degrees of freedom only. The two observables $\M$ and $I_{\alpha\beta}$ provide two quantitatively different measures of spontaneous T-breaking in the orbital degrees of freedom of a given material. On the experimental side, $I_{\alpha\beta}$ is naturally endowed with core chemical specificity, at the root of its great success; while instead only the total $\M$ value is experimentally accessible.

Condensed matter physics adopts Born-von-K\`arm\`an  periodic boundary conditions (PBCs) for either crystalline or disordered materials; but PBCs are incompatible with the presence of a generic magnetic field. Because of the same reasons, the angular momentum $L_z$ is a ``forbidden'' operator. The theory of orbital magnetization was established in 2006 \cite{rap130,Vanderbilt}

Here I thoroughly investigate analogies and differences between the two observables, also providing three significant advances. (i) I give a microscopic expression for $I_{\alpha\beta}$, gauge-invariant in form, for any crystalline material (either metal or insulator) within band-structure theory; I also show that, in a Chern insulator, $I_{\alpha\beta}$ is not affected by the topologically protected edge states (while $\M$ is affected) (ii) I show that $I_{\alpha\beta}$ is a {\it local} observable, in full analogy to $\M$ \cite{rap148,rap150}; it must be stresssed that other geometrical observables are strongly nonlocal (most notably electrical polarization \cite{rap148}). (iii) I go beyond band-structure theory, and I show that even in a correlated many-electron system $I_{\alpha\beta}$ is a geometrical ground-state observable, although in a different Hilbert space. When PBCs are abandoned, the very same geometrical formula for $I_{\alpha\beta}$ goes seamlessly into the center-of-mass angular momentum formula of Ref. \cite{Kunes00} (which only makes sense for a bounded crystallite).

\section{Theory}

Shortly after the theory of orbital magnetization was fully established \cite{rap130}, Souza and Vanderbilt derived an explicit expression for $I_{\alpha\beta}$  in the special case of a topologically trivial insulator \cite{Souza08}. It was shown that both $\M$ and $I_{\alpha\beta}$ are geometrical properties of the electronic ground state; an explicit expression for their difference was also provided.

The XMCD sum rule concerns the frequency integral of the imaginary part of the antisymmetric term in the conductivity tensor \[ I_{\alpha\beta} = \mbox{Im } \int_0^\infty d\omega \; \sigma^{(-)}_{\alpha\beta}(\omega) ; \] a kind of fluctuation-dissipation theorem relates $I_{\alpha\beta}$ to a ground-state property, both for a bounded sample (e.g. a crystallite) within the so-called ``open'' boundary conditions (OBCs), and for an unbounded solid within PBCs. In both frameworks, all ground-state properties---at the independent-particle level---can be expressed in terms of the relevant ground-state projector $\CP$.

In the OBCs case the projector (per spin channel) is: \[ \CP = \sum_{\epsilon_j \leq \mu} \ket{\varphi_j}\bra{\varphi_j} , \label{obc} \] where $\mu$ is the Fermi level and $\ket{\varphi_j}$ are the single particle eigenstates of the Hamiltonian $\CH$ with eigenvalues $\epsilon_j$. In the band-structure case the projector is instead \[ \CP = \vc \intk \sum_{\epsilon_{j\k} \leq \mu}  \ket{\psi_{j\k}} \bra{\psi_{j\k}} , \label{bloch1} \] where BZ is the Brillouin zone, $\ket{\psi_{j\k}}$ are the Bloch orbitals normalized to one in the crystal cell of volume $\vc$, $\epsilon_{j\k}$ are the band energies, and $d$ is the dimension; $\vc$ must be understood as the area for $d=2$. The reason for adopting the same symbol $\CP$ in \eqs{obc}{bloch1} lies in the ``nearsightedness'' principle \cite{Kohn96}. If one evaluates $\CP$ from \equ{obc} for a large bounded crystallite, and then further projects this $\CP$ onto the inner region of the crystallite, the result asymptotically converges to the $\CP$ value provided by \equ{bloch1} for the same material; the convergence is exponential in insulators and power law in metals.

We start with a bounded sample within OBCs: the sum rule for $I_{\alpha\beta}$ has a relatively straighforward expression \cite{Souza08}; a tedious calculation (see the Appendix) shows that it can be equivalently expressed as: \[ I_{\alpha\beta} = - \frac{i \pi e^2}{2 \hbar^2 V} \mbox{Tr } \{ (\CH - \mu) \, [\,[r_\alpha, \CP] , [r_\beta, \CP] \,] \}  . \label{main}\]  The virtue of this expression becomes clear when switching to PBCs and band-structure theory. In the latter case $\CP$ is lattice periodical: \[ \me{\r}{\CP}{\r'} = \me{\r+\R}{\CP}{\r'+\R}, \] where $\R$ is a lattice vector. The position $\r$ is instead a forbidden operator, incompatible with PBCs \cite{rap100}; notwithstanding, the commutator $[\r, \CP]$ is an honest lattice-periodical operator (like $\CH$ and $\CP$), hence \equ{main} can be adopted as it is: \[ I_{\alpha\beta} = - \frac{i \pi e^2}{2 \hbar^2 \vc} \mbox{Tr$_{\rm cell}$ } \{ (\CH - \mu) \, [\,[r_\alpha, \CP] , [r_\beta, \CP] \,] \} . \label{main2} \] Therein, $\CH$ is the periodic Hamiltonian and $\CP$ is given by \equ{bloch1}. 

The next step is to express \equ{main2} as a Fermi volume integral (BZ integral in the insulating case), as customary for any intensive ground-state observable; its geometrical nature will also perspicuously appear. We cast \equ{bloch1} in terms of the periodic Bloch orbitals $\ket{u_{j\k}}$ = $\emi{\k \cdot \r} \ket{\psi_{j\k}}$, eigenvectors of $\CH_\k = \emi{\k \cdot \r} \CH \ei{\k \cdot \r}$: \bea \me{\r}{\CP}{\r'} &=& \vc \intk \ei{\k \cdot (\r -\r')} \me{\r}{\CP_\k}{\r'} \label{CP1} \\ \CP_{\k} &=&  \sum_{\epsilon_{j\k} \leq \mu}  \ket{u_{j\q}} \bra{u_{j\q}} . \label{CP2} \eea 

In order to establish a differential geometry in the space of the $\ket{u_{j\k}}$ state vectors we choose a gauge which makes the $\ket{u_{j\k}}$ smooth (i.e. $C^\infty$) throughout the whole BZ; this is always possible, even in topologically nontrivial materials \cite{mona}. The integrand in \equ{CP1} is periodical in $\k$, hence the BZ integral of its $\k$-derivative vanishes: \[ 0 = i(\r - \r') \me{\r}{\CP}{\r'} + \vc \intk \ei{\k \cdot (\r -\r')} \me{\r}{\dkk \CP_\k}{\r'} \] \[ i [\r, \CP] = - \vc \intk \ei{\k\cdot\r} \dkk \CP_\k \emi{\k\cdot\r} .\] We are now ready to replace this into \equ{main2}, together with \[ \CH = \vc \intk \ei{\k\cdot\r} \CH_\k \emi{\k\cdot\r} . \] The three reciprocal-space integrals in the product contract to one (see the Appendix), and we arrive at \[ I_{\alpha\beta} = \frac{i \pi e^2}{2 \hbar^2} \intk \mbox{Tr$_{\rm cell}$} \{ (\CH_\k - \mu)  [\da \CP_\k,\db \CP_\k ] \} .  \label{major} \] \equ{major} is one of the major results of the present work; it applies on the same ground to trivial insulators, topological insulators, and metals. In the metallic case, the $\k$-derivative of $\CP_\k$ includes a $\delta$-like singularity at the Fermi level, which is annihilated by antisymmetrization. When specialized to the so-called ``Hamiltonian gauge'' \cite{Vanderbilt} \equ{major} yields the same formula as in Ref. \cite{Souza08} (derived therein in a completely different way for trivial insulators only).

As said above, orbital magnetization $\M$ probes T-breaking differently than $I_{\alpha\beta}$; for comparison I am going to display the analogous expression for $\M$ \cite{rap130,Vanderbilt} when expressed within the present formalism. Preliminarly, we observe that custom dictates the adoption of the field ${\bf H}$ in the free energy; 
instead, because of the reasons well explained in Ref. \cite{Griffiths-e},
first-principle theory alternatively adopts the more fundamental field $\B$. 
The (by now classic) orbital magnetization expression, formulated in the Hamiltonian gauge  \cite{rap130,Vanderbilt}, can be shown to be equivalent to the gauge-invariant expression \cite{rap} \[ \M =\frac{i e}{2 \hbar c} \intk \mbox{Tr$_{\rm cell}$} \{ |\CH_\k -\mu| (\dkk \CP_\k) \times (\dkk \CP_\k) \} , \label{M} \] where $|\CH_\k -\mu| = (\CH_\k -\mu) ({\cal I} - 2 \CP_\k)$ is the operator which acts as $(\mu-\CH_k)$ on the occupied states, and as $(\CH_\k -\mu)$ on the unoccupied ones. Comparing to the XMCD sum rule, \equ{major}, the key difference is the presence/absence of the modulus of the operator  $(\CH_\k -\mu)$ (besides an obvious prefactor). 

We stress that $\CP_\k$ is gauge-invariant in the generalized sense, i.e. by arbitrary mixing of the occupied $\ket{u_{j\k}}$ orbitals at a given $\k$ \cite{Vanderbilt}; therefore \eqs{major}{M} are gauge-invariant in form. Their geometrical integrands are simple modifications of the Berry curvature, whose known expression is: \[\Omega_{\alpha\beta}(\k) = i \mbox{Tr$_{\rm cell}$} \{ \CP_\k  [\da \CP_\k,\db \CP_\k ] \} . \] Therein the band projector $\CP_\k$ is the sole ingredient, while the Hamiltonian enters \eqs{major}{M} as well; it enters both formul\ae\ via 
$(\CH_\k -\mu)$ for aesthetical reasons, i.e. to explicitly display their invariance by translation of the energy zero.  Nonetheless, the $\mu$-dependence can be dropped in $I_{\alpha\beta}$ since the $\mu$-dependent term is annihilated after antisymmetrization. In this regard, $\M$ is different: it is $\mu$-independent in trivial insulators, while it actually depends on $\mu$ (in the gap) in Chern insulators \cite{rap151}. Remarkably, this means that---in a bounded sample---the topologically protected boundary states contribute to $\M$, but not to $I_{\alpha\beta}$.

Finally I point out another major appealing feature of \eqs{main2}{major}: both show the {\it locality} of the observable. If we define the (real) lattice-periodical function $\CX_{\alpha\beta}(\r)$ as \[ \CX_{\alpha\beta}(\r) = \frac{\pi e^2}{2 \hbar^2} \mbox{Im } \me{\r}{(\CH - \mu) [\, [r_\alpha, \CP] , [ r_\beta, \CP ] \;]}{\r} \]  the sum rule reads  \[I_{\alpha\beta} = \frac{1}{\vc} \int_{\rm cell} d\r \; \CX_{\alpha\beta}(\r) . \] The function $\CX_{\alpha\beta}(\r)$ assumes then the meaning of a XMCD ``density'' and provides spatial resolution, although such resolution is not expected to be microscopic, only macroscopic. Previous experience with other local geometrical observables, mostly based on model Hamiltonians,  has shown---when dealing e.g. with an heterostructure---that the ``density'' separates the contributions from the different regions of the sample \cite{rap146,rap153,rap156}. Obviously, the $\k$-space formula in the supercell formalism lacks spatial resolution even at the macroscopic level.

The concept of orbital magnetization density has been proposed and validated in Refs. \cite{rap148,rap151,rap150}; its macroscopic average coincides by definition with (minus) the $\B$-derivative of the orbital free-energy density (spin magnetization is obviously local as well). As said in the Introduction, the {\it apparently} analogous case of electrical polarization is instead strongly nonlocal: no ``density'' can be defined  \cite{Vanderbilt,rap148}.

\section{Contribution from an isolated flat band}

Whenever the band spectrum comprises an isolated band, its contribution to both observables can be evaluated by including that band only in the ground-state projector. This parallels the spectral selectivity naturally provided by the sum rule, and largely exploited by XMCD experimentalists. The single-band expressions for the two observables are, in case of full occupancy \cite{rap130,Souza08,Vanderbilt}:  \[ I_{\alpha\beta} = \frac{\pi e^2}{\hbar^2} \mbox{Im}  \intk \me{\da u_\k}{(\CH_\k - \epsilon_\k)}{\db u_\k}, \label{single1} \]\[ M_\gamma =  \frac{e }{2 \hbar c} \varepsilon_{\gamma\alpha\beta} \;\mbox{Im} \intk \me{\da u_\k}{(\CH_\k + \epsilon_\k)}{\db u_\k}, \label{single2}\] where summation over repeated indices is implicit, and $\varepsilon_{\gamma\alpha\beta}$ is the antisymmetric tensor (in the case of $\M$ the argument of ``Im'' is purely imaginary after antisymmetrization). For a flat core band the difference between the two integrals is a constant times the Chern invariant, which in turn vanishes by assumption. 

We remind that in the general case the orbital moment per cell as such is an ill defined quantity: both $\M$ and $I_{\alpha\beta}$ have the primary meaning of a free-energy derivative. Only in the extreme case of an isolated flat band they both can be alternatively expressed as the core orbital moment per cell.

Switching to vector notations, \equ{single2} becomes \[ \M = - \frac{i e}{2\hbar c}  \intk \me{\dkk u_\k}{\CH_\k \times}{\dkk u_\k}.  \label{single} \] If $\varphi(\r)$ is the core orbital in the Schr\"odinger representation, the zero-hopping assumption yields: \[ \ev{\r|u_\k} = \sum_\R \ei{\k \cdot(\R-\r)} \varphi(\r-\R), \]\[ \ev{\r|\dkk u_\k} = i \sum_\R (\R-\r) \ei{\k \cdot(\R-\r)} \varphi(\r-\R) , \] where $\R$ are the lattice vectors. Exploiting the zero-overlap assumption, and the fact that $\CH_\k$ commutes with $\ei{\k\cdot\R}$ we get \bea \me{\dkk u_\k}{ \CH_\k \times}{\dkk u_\k} &=& \int_{\rm cell} d \r \sum_\R \varphi^*(\R-\r) \ei{\k \cdot \r}  \CH_\k  \nn &\times& \emi{\k\cdot\r} (\R-\r)\varphi(\r-\R) .\eea Since $\CH_\k = \emi{\k \cdot \r} \CH \ei{\k\cdot\r}$, the above expression is $\k$-independent. Then  \[ \M = - \frac{i e}{2\hbar c \vc}  \me{\dkk u_\k}{ \CH_\k \times}{\dkk u_\k}  =  \frac{i e}{2\hbar c \vc} \me{\varphi}{\CH \times \r}{\varphi} . \] Finally using $\r \times \r =0$ and ${\bf v} = i[\CH,\r]/\hbar$, we obtain \[ \M  = \frac{i e}{2\hbar c \vc} \me{\varphi}{[\CH,\r] \times \r}{\varphi} = - \frac{e}{2 c \vc} \me{\varphi}{\r \times {\bf v} }{\varphi} , \label{cm} \] thus retrieving the magnetic analogue of the popular Clausius-Mossotti model polarization in electrostatics.
Such an expression makes sense only because the electron distribution is confined in disconnected core regions, separated by empty interstitial space; therefore the unbounded nature of $\r$ makes no harm \cite{rap100}. 

In most cases---such as for $d$ levels in transition metals---the band is not flat, not isolated, and/or composite; nonetheless most of XMCD literature \cite{Thole92,Wu93,Benoist00} assumes that $I_{\alpha\beta}$ provides an approximate measure of orbital magnetization. The identification of the two integrals in \eqs{single1}{single2} is not justified, not even approximately: the Berry curvature is known to be a weird function of $\k$ (see e.g. Fig. 5.8 in Ref. \cite{Vanderbilt}). 

\section{Correlated ground state}

At this point we abandon band-structure theory altogether and we switch dealing with an interacting system. The formula was provided by Kunes and Oppeneer in 2000 \cite{Kunes00} for a bounded crystallite (or even a noncrystalline sample); it cannot be adopted within PBCs, given that one of its entries is the position operator $\r$, forbidden within PBCs \cite{rap100}. Here we provide the many-body formula in terms of a many-body ground state, and we show that even in this case $I_{\alpha\beta}$ is a geometrical observable, although in a different Hilbert space. For the sake of simplicity we deal with the simple case where a purely orbital Hamiltonian can be established for each spin channel (same as above for the band-structure case). 

We consider, following a milestone paper by W. Kohn \cite{Kohn64}, a system of $N$ interacting particles in a cubic box of volume $L^3$, and the family of many-body Hamiltonians parametrized by the parameter $\kk$: \[ \hat{H}_{\kk} = \frac{1}{2m} \sum_{i-1}^N \left[\p_i + \frac{e}{c} {\bf A}(\r_i) + \hbar \kk \right]^2 + \hat{V}, \label{kohn} \] where $\hat{V}$ includes one-body and two-body potentials. The vector potential ${\bf A}(\r)$ summarizes all T-breaking terms, as e.g. those due to spin-orbit coupling to a background of local moments.
We adopt Born-von-K\`arm\`an PBCs over each
electron coordinate $\r_i$ independently, whose Cartesian components
$r_{i,\alpha}$ are then equivalent to the angles $2\pi r_{i,\alpha}/L$. The
potential $\hat{V}$ and the vector potential ${\bf A}(\r)$ enjoy the same periodicity. The vector $\kk$, having the dimensions of an inverse length, is called ``flux'' or ``twist''. Setting $\kk \neq 0$ amounts to a gauge transformation; since PBCs violate gauge-invariance, the eigenvectors $\ket{\Psi_{n\kk}}$ and the eigenvalues $E_{n\kk}$ have a nontrivial $\kk$-dependence \cite{Kohn64}.

In order to simplify notations we define $\hat{H}_{0} \equiv \hat{H}$, $\ket{\Psi_{n0}} \equiv \ket{\Psi_{n}}$ , $E_{n0} \equiv E_n$, and $\omega_{0n} = (E_n - E_0)/\hbar$. If we further define the $\kk =0$ many-body velocity operator as \[ \hat{\bf v} = \frac{1}{m} \sum_{i-1}^N \left[\p_i + \frac{e}{c} {\bf A}(\r_i) \right] = \frac{1}{\hbar} \partial_{\kk} \hat{H} , \] the Kubo formula for the imaginary part of the conductivity straightforwardly yields the sum rule:  
 \[ I_{\alpha\beta} = \frac{\pi e^2}{\hbar L^3} \mbox{Im}\sum_{n\neq 0} \frac{ \me{\Psi_0}{{\hat v}_\alpha}{\Psi_n}\me{\Psi_n}{{\hat v}_\beta}{\Psi_0}}{\omega_{0n}} ; \label{corr}\] the task is now to convert this expression into a ground-state property.

The many-body analogue of the $\k \cdot \p$ expansion yields \[ \ket{\partial_{\kk} \Psi_0} = - \sum_{n\neq 0} \ket{\Psi_n} \frac{\me{\Psi_n}{\hat{\bf v}}{\Psi_0}}{\omega_{n0}} \] \[ (\hat{H} - E_0) \ket{\partial_{\kk} \Psi_0} = - \sum_{n\neq 0} \ket{\Psi_n} \me{\Psi_n}{\hat{\bf v}}{\Psi_0} \] By comparing these expression to \equ{corr} we find our final geometrical formula
\[ I_{\alpha\beta} = \frac{\pi e^2}{\hbar^2 L^3} \mbox{Im } \me{\partial_{\kappa_\alpha} \Psi_0 }{(\hat{H} - E_0)}{\partial_{\kappa_\beta} \Psi_0} .\] The correspondence with the single-band formula of \equ{single1} is self evident; a gauge-invariant expression can be easily obtained from the ground-state projector $\hat{P}_{\kk} = \ket{\Psi_{\kk}}\bra{\Psi_{\kk}} $.
At variance with the band-structure case, the many-body formula for $I_{\alpha\beta}$ unfortunately cannot be compared with a corresponding formula for $\M$. To this day such a formula does not exist: the orbital magnetization of a corrrelated many-body wavefunction within PBCs is currently an open (and challenging) problem. The corresponding problem for electrical polarization has been solved long ago \cite{rap100}.

From now on we adopt the same Kohn Hamiltonian of \equ{kohn}, but we switch form PBCs to OBCs, thus addressing a crystallite (or any bounded sample). In this case the flux $\kk$ can be easily ``gauged away'': the energies are $\kk$-independent, while the state vectors are \[ \ket{\Psi_{n\kk}} = \emi{\kk \cdot \hat{\r}} \ket{\Psi_{n}}, \qquad \mbox{OBCs} , \] where $\hat{\r} = \sum_i \r_i$ is the many-body position, a trivial multiplicative operator within OBCs. Therefore \[ \ket{\partial_{\kk} \Psi_0} = - i \hat{\r} \ket{\Psi_0} , \] \bea I_{\alpha\beta} &=&  \frac{\pi e^2}{\hbar^2 L^3} \mbox{Im } \me{\Psi_0 }{\hat{r}_\alpha(\hat{H} - E_0)\hat{r}_\beta}{\Psi_0} \nn &=&  - \frac{i \pi e^2}{2\hbar^2 L^3} \mbox{Im } \me{\Psi_0 }{\hat{r}_\alpha [\hat{H},\hat{r}_\beta]}{\Psi_0} \nn &=& - \frac{\pi e^2}{2\hbar L^3}  \me{\Psi_0 }{(\hat{r}_\alpha\hat{v}_\beta -\hat{r}_\beta\hat{v}_\alpha)}{\Psi_0} . \label{kunes} \eea Not surprisingly, this is identical to the center-of mass angular momentum formula, first derived in Ref. \cite{Kunes00}.

By expanding the many-body operators $\hat{\r}$ and $\hat{{\bf v}}$, \equ{kunes} is seen to be proportional to the ground-state expectation value of $\sum_{ii'} \r_i \times {\bf v}_{i'}$, while the orbital moment of a bounded sample is proportional to the ground-state expectation value of $\sum_{i} \r_i \times {\bf v}_{i}$. The two coincide only in the single-electron case, consistently with \equ{cm} above: the case of an isolated flat band amounts in fact to considering one electron per cell (and per spin channel).

\section{Conclusions}
 
I have presented a microscopic expression for the XMCD sum rule $I_{\alpha\beta}$ within band-structure theory, and compared it to the analogous formula for orbital magnetization $\M$. Both geometrical formul\ae\ are gauge-invariant in form and very general: they address trivial insulators, topological insulators, and metals on the same footing. In a Chern insulator the topologically protected edge states contribute to $\M$; I prove here that they do not contribute to $I_{\alpha\beta}$.
I have also proved the locality of $I_{\alpha\beta}$, i.e. its spatial-resolution capability ($\M$ is known to be local \cite{rap148,rap150}). In agreement with common wisdom, the present formulation confirms that the contributions to $I_{\alpha\beta}$ and to $\M$ by an isolated flat band of nonoverlapping core states coincide, and also coincide with the intuitive---Clausius-Mossotti inspired---definition of $\M$ as the core orbital moment over the cell volume. I have also presented the many-body version of $I_{\alpha\beta}$, which is a geometrical ground-state observable as well, although in a different Hilbert space.

It is difficult to say anything about to which extent the two observables could provide similar results in some classes  of materials (and perhaps very different in others), for the good reason that few first-principle calculations of $\M$ exist \cite{Ceresoli10,Hanke}, and none of the sum rule (to the best of my knowledge). A further drawback is that ``first-principle'' is regarded as a synonymous of density-functional theory, whose accuracy in addressing T-breaking properties and materials is known to be poor \cite{Ceresoli10}. 

With respect to magnetization measurements, the XMCD sum rule $I_{\alpha\beta}$ has two great virtues, at the root of its enormous success. The first one is that $I_{\alpha\beta}$ is a purely orbital property, while in magnetization measurements the spin contribution has to be subtracted; further complications arise from shape issues and demagnetization coefficients.  The second major virtue is that the frequency integral can be (and routinely is) performed over selected spectral ranges, thus isolating the T-breaking contributions from selected orbital shells, and providing element and/or site specific information. The XMCD sum rule $I_{\alpha\beta}$ is an outstanding ground-state observable per se; there is no compelling reason for identifying it (even approximately) with some form of orbital magnetization.

\section*{Acknowledgments} 
 
Useful discussions with M. Altarelli and I. Souza are gratefully acknowledged. Work supported by the ONR Grant No. No. N00014-17-1-2803.

\section*{Appendix}

\subsection*{XCMD sum rule in a bounded sample}

Besides $\CP$ we need here its complementary $\CQ = {\cal I} - \CP$: \[ \CQ = \sum_{\epsilon_j > \mu} \ket{\varphi_j}\bra{\varphi_j} . \]  Following Ref. \onlinecite{Souza08} the XCMD sum rule reads 
\bea I_{\alpha\beta} &=&\frac{\pi e^2}{2 \hbar V} \mbox{Tr } \{ \CP r_\alpha \CQ v_\beta\} - (\alpha \leftrightarrow \beta) \nn &\doteq& \frac{\pi e^2}{2 \hbar V} \mbox{Tr } \{ \CP r_\alpha \CQ v_\beta\} \nn &\doteq& \frac{i \pi e^2}{2 \hbar^2 V} \mbox{Tr } \{ \CP r_\alpha \CQ \, [\CH,r_\beta] \}, \eea where the symbol ``$\doteq$'' means ` equal after antisymmetrization, and $v_\beta= i [\CH,r_\beta] / \hbar$ has been used. We then switch to a $\CP$-only formula: \bea \mbox{Tr } \{ \CP r_\alpha \CQ \, [\CH,r_\beta] \} &\doteq&  \mbox{Tr } \{ \CP r_\alpha \CQ \CH r_\beta \}  - \mbox{Tr } \{ \CP r_\alpha \CQ r_\beta \CH \}  \nn &\doteq& \mbox{Tr } \{ \CP r_\alpha \CH r_\beta \}  - \mbox{Tr } \{ \CP r_\alpha \CP \CH r_\beta \} \nn &+& \mbox{Tr } \{ \CP r_\alpha \CP r_\beta \CH \} , \label{P} \eea where a symmetric term has been dropped.
Using then the cyclic invariance of the trace, and the fact that $\CH$ commutes with $\CP$, \equ{P}  is recast as \bea \mbox{Tr } \{ \CP r_\alpha \CQ \, [\CH,r_\beta] \} &\doteq&   \mbox{Tr } \{ \CH r_\beta \CP r_\alpha  \}  - \mbox{Tr } \{ \CH \CP r_\beta \CP r_\alpha \} \nn &+& \mbox{Tr } \{ \CH \CP r_\alpha \CP r_\beta \} \nn &\doteq& \mbox{Tr } \{ \CH (2 \CP - {\cal I}) r_\alpha \CP r_\beta  \} . \eea It is then easy to verify that \[ \mbox{Tr } \{ \CH (2 \CP - {\cal I}) r_\alpha \CP r_\beta  \} \doteq - \mbox{Tr } \{ \CH [r_\alpha, \CP] \, [r_\beta, \CP]  \} , \] where again a symmetric term has been dropped. 

We thus arrive at: \[ I_{\alpha\beta} = - \frac{i \pi e^2}{2 \hbar^2 V} \mbox{Tr } \{ \CH \, [\,[r_\alpha, \CP] , [r_\beta, \CP] \,] \}  . \] One further aesthetical improvement is desirable. We observe that any bulk observable (except the energy itself) must be invariant by translation of the energy zero, hence must be a function of $\CH - \mu$; and in fact $\CP$ can be formally written as $\CP=\theta(\mu - \CH)$, where $\theta$ is the step function. We therefore we rewrite \[ I_{\alpha\beta} = - \frac{i \pi e^2}{2 \hbar^2 V} \mbox{Tr } \{ (\CH - \mu) \, [\,[r_\alpha, \CP] , [r_\beta, \CP] \,] \}  . \label{final} \] 

\subsection*{Products of lattice-periodical operators}

We start with a simple lemma: if $f(\r)$ is a lattice-periodical function, then its Fourier transform is \[ \int \ei{\k\cdot\r} f(\r) = \frac{(2\pi)^d}{\vc} \delta(\k) \intrc f(\r) .\]
Any two lattice periodical operators can be written in Schr\"odinger representation as \bea \me{\r}{\CA}{\r'} &=& \vc \intk \ei{\k \cdot (\r -\r')} \me{\r}{\CA_\k}{\r'} , \nn \CA_k &=& \sum_{jj'} \ket{u_{j\k}} \me{\psi_{j\k}}{\CA}{\psi_{j'\k}} \bra{u_{j'\k}} . \eea  \bea \me{\r}{\CB}{\r'} &=& \vc \intk \ei{\k \cdot (\r -\r')} \me{\r}{\CB_\k}{\r'} , \nn \CB_k &=& \sum_{jj'} \ket{u_{j\k}} \me{\psi_{j\k}}{\CB}{\psi_{j'\k}} \bra{u_{j'\k}} . \eea 
Their products obtain as \bea \me{\r}{\CA\CB}{\r'} &=& \int d\r'' \me{\r}{\CA}{\r''} \me{\r''}{\CB}{\r'} \\  \me{\r}{\CA_\k\CB_{\k'}}{\r'}, &=& \int_{\rm cell} d\r'' \me{\r}{\CA_\k}{\r''} \me{\r''}{\CB_{\k'}}{\r'} . \eea
Notice that the integral in $d\r''$ is over ${\mathbb R}^d$ in the case of $\CA\CB$, and over the crystal cell in the case of  $\CA_\k\CB_{\k'}$; notice also that $\CA_\k$ and $\CB_{\k'}$ are lattice-periodical in $\r$ and $\r'$ {\it independently}. 

We address now the diagonal term $\me{\r}{\CA\CB}{\r}$: a lattice-periodical function. Its expression is: \bea \me{\r}{\CA\CB}{\r} &=& \vc^2 \intk \int \frac{d \k'}{(2\pi)^d} \ei{(\k - \k') \cdot \r} \times \nn &\times& \int d\r'' \ei{(\k'-\k) \cdot \r''} \me{\r}{\CA_\k}{\r''} \me{\r''}{\CB_{\k'}}{\r} . \nonumber
\eea 
By exploiting the above lemma in the last line one gets \bea \me{\r}{\CA\CB}{\r} &=& \vc^2 \intk \int \frac{d \k'}{(2\pi)^d} \ei{(\k - \k') \cdot \r} \times \nn &\times& \frac{(2\pi)^d}{\vc} \delta(\k'-\k) \int_{\rm cell} d\r'' \me{\r}{\CA_\k}{\r''} \me{\r''}{\CB_{\k'}}{\r} \nn &=& \vc \intk \me{\r}{\CA_\k\CB_\k}{\r} \nonumber. \eea This process is associative, and can be extended to the product of three operators: \[ \me{\r}{\CA\CB\CC}{\r} = \vc \intk \me{\r}{\CA_\k\CB_\k\CC_\k}{\r} . \]

In the main text we identify $\CA$, $\CB$, and $\CC$ with the lattice-periodical Hermitian  operators $(\CH - \mu)$, $i[r_\alpha,\CP]$, and $i[r_\beta,\CP]$, respectively. Correspondently, they enter the Fermi volume integral as $(\CH_\k - \mu)$, $-\da \CP_\k$, and $-\db \CP_\k$.

 %\bibliography{$HOME/inputs/huge_bib}
%\bibliography{$HOME/inputs/huge_bib,add_bib}

\begin{thebibliography}{10}

\bibitem{Thole92}
{ B. T. Thole, P. Carra, F. Sette, and G. van der Laan, Phys. Rev. Lett. {\bf
  68}, 1943 (1992)}.

\bibitem{Kunes00}
{ J. Kunes and P. M. Oppeneer, Phys. Rev. B {\bf 61}, 15774 (2000)}.

\bibitem{rap130}
{ D. Ceresoli, T. Thonhauser, D. Vanderbilt, and R. Resta, Phys. Rev. B {\bf
  74}, 024408 (2006)}.

\bibitem{Vanderbilt}
{ D. Vanderbilt, {\it Berry Phases in Electronic Structure Theory} (Cambridge
  University Press, Cambridge, 2018)}.

\bibitem{rap148}
{ R. Bianco and R. Resta, Phys. Rev. Lett. {\bf 110}, 087202 (2013)}.

\bibitem{rap150}
{ A. Marrazzo and R. Resta, Phys. Rev. Lett. {\bf 116}, 137201 (2016)}.

\bibitem{Souza08}
{ I. Souza and D. Vanderbilt, Phys. Rev. B {\bf 77}, 054438 (2008)}.

\bibitem{Kohn96}
{ W. Kohn, Phys. Rev. Lett. {\bf 76}, 3168 (1996)}.

\bibitem{rap100}
{ R. Resta, Phys. Rev. Lett. {\bf 80}, 1800 (1998)}.

\bibitem{mona}
{ When the Chern number is nonzero the smooth Bloch orbitals cannot obey the
  so-called periodic gauge \cite{Vanderbilt}; conversely, if chosen as
  periodical in $\k$ they cannot be smooth}.

\bibitem{Griffiths-e}
{ D. J. Griffiths, {\it Introduction to Electrodynamics, 3rd Ed.}
  (Prentice-Hall, 1999)}.

\bibitem{rap}
{ R. Resta, https://arxiv.org/abs/1703.00712 (rejected by Phys. Rev. Lett.)}.

\bibitem{rap151}
{ R. Bianco and R. Resta, Phys. Rev. B {\bf 93}, 174417 (2016)}.

\bibitem{rap146}
{ R. Bianco and R. Resta, Phys. Rev. B {\bf 84}, 241106(R) (2011)}.

\bibitem{rap153}
{ A. Marrazzo and R. Resta, Phys. Rev. B {\bf 95}, 121114(R) (2017)}.

\bibitem{rap156}
{ A. Marrazzo and R. Resta, Phys. Rev. Lett. {\bf 122}, 166602 (2019)}.

\bibitem{Wu93}
{ R. Wu, D. Wang, and A. J. Freeman, Phys. Rev. Lett. {\bf 71}, 3581 (1993)}.

\bibitem{Benoist00}
{ R. Benoist, P. Carra, and O. K. Andersen, Eur. Phys. J. B {\bf 18}, 193
  (2000)}.

\bibitem{Kohn64}
{ W. Kohn, Phys. Rev. {\bf 133}, {A171} (1964)}.

\bibitem{Ceresoli10}
{ D. Ceresoli, U. Gerstmann, A. P. Seitsonen, and F. Mauri, Phys. Rev. B {\bf
  81}, 060409(R) (2010)}.

\bibitem{Hanke}
{ See J.-P. Hanke {\it et al.}, Phys Rev B {\bf 94}, 121114 (2016) and
  references quoted therein}.

\end{thebibliography}
%\bibliographystyle{unsrt}

\end{document}